# Hybrid complex network topologies are preferred for component-subscription in large-scale data-centres

Ilango Sriram and Dave Cliff

Department of Computer Science, University of Bristol
Bristol, UK
{ilango, dc}@cs.bris.ac.uk

**Abstract.** We report on experiments exploring the interplay between the topology of the complex network of dependent components in a large-scale data-centre, and the robustness and scaling properties of that data-centre. In a previous paper [1] we used the *SPECI* large-scale data-centre simulator [2] to compare the robustness and scaling characteristics of data-centres whose dependent components are connected via Strogatz-Watts small-world (SW) networks [3], versus those organized as Barabasi-Albert scale-free (SF) networks [4], and found significant differences. In this paper, we present results from using the Klemm-Eguiluz (KE) construction method [5] to generate complex network topologies for data-centre component dependencies. The KE model has a control parameter $\mu \in [0,1] \in \mathbf{R}$ that determines whether the networks generated are SW ($0<\mu<<1$) or SF ($\mu=1$) or a "hybrid" network topology part-way between SW and SF ($0<\mu<1$). We find that the best scores for system-level performance metrics of the simulated data-centres are given by "hybrid" values of $\mu$ significantly different from pure-SW or pure-SF.

## 1 Introduction

Modern ultra-large-scale data-centres appear, from the outside at least, to be highly regimented and regularly-structured engineering artefacts. Aisle after aisle of racks, each rack being a vertical frame housing a number of chasses, each chassis housing a regular arrangement of thin computer mother-board units: the "blade-servers" that make up the data-centre's computing infrastructure. In this paper, we show that *hybrid* complex network topologies of dependencies between the computing components that make up a data-centre service may bring benefits. We emphasise the hybrid nature of the complex network topologies here because the results published in this paper demonstrate that "non-standard" network constructions are best: these topologies are non-standard in the sense that they are neither purely scale-free (SF) nor purely small-world (SW), but rather part-way between the two.

   Data centres are becoming ever-larger as the operators seek to maximise the benefits from economies of scale. With these increases in size comes a growth in system complexity, which is usually problematic. The growth in complexity manifests itself in two primary ways. The first is that many conventional management techniques that work well when controlling a relatively small number of data-centre nodes scale much worse than linearly and hence become impracticable when the number of nodes under control increases by two or three orders of magnitude. The second is that the very large number of individual independent hardware components in modern data centres means that, even with very reliable components, at any one time it is reasonable to expect there always to be one or more significant component failures (so-called "normal failure"). Despite this, guaranteed levels of performance and dependability must be maintained. For an extended discussion of the issues that arise in the design of warehouse-scale data-centres, see [6].

Predictive computer simulations are used in almost all of current engineering practice, to evaluate possible designs before they go into production. Simulation studies allow for the rapid exploration and evaluation of design alternatives, and can help to avoid costly mistakes. In microelectronics e.g., the well-known *SPICE* circuit-simulation system [7] has long allowed large-scale, highly complex designs to be evaluated, verified, and validated in simulation before the expensive final stage of physical fabrication. Despite this well-established tradition of computational modelling and simulation tools being used in other engineering domains, there are currently no comparable tools for simulating cloud-scale computing data-centres. The lack of such tools prevents the application of rigorous formal methods for testing and verifying designs before they go into production. Put bluntly, at the leading edge of data-centre design and implementation, current practice is much more art than science, and this imprecision can lead to costly errors.

As a first step in meeting this need, we have developed *SPECI* (Simulation Program for Elastic Cloud Infrastructures). Clearly, it would require many person-years of effort to bring *SPECI* up to the comprehensive level of *SPICE* or of commercial CFD tools. We are currently exploring the possibility of open-sourcing *SPECI* in the hope that a community of contributors then helps refine and extend it.

The first paper discussing *SPECI* [2] gave details of its rationale and design architecture that will not be repeated here. In that first paper, results were presented from simulation experiments that had been suggested by our industrial sponsor, Hewlett-Packard Laboratories. The specific area of inquiry is large-scale data-centre middleware component-status subscription-update policies. The status of data-centre components may change as they fail, or as policies are updated. Within the data-centre, there will be components that work together and need to know the status of other components via "subscriptions" to status-updates from those components. In [2] we used a first-approximation assumption that such subscriptions are distributed randomly across the data centre. That is, the connectivity of the network of subscription dependencies within the data-centre was, formally, a random graph. In [1], we explored the effects of introducing more realistic constraints to the structure of the internal network of subscriptions. We explored the effects of making the data-centre's subscription network have a regular lattice-like structure, and also the effects when the networks had complex semi-random structures resulting from parameterised network generation functions that create small-world [3] and scale-free [4] networks. We showed that for distributed middleware topologies, varying the structure and distribution of tasks carried out in the data centre can significantly influence the performance overhead imposed by the middleware.

In this paper we inspect the use of hybrid complex network topologies as basis for the structure and distribution of tasks carried out in the data centre. We use the

Klemm-Eguiliz (KE) construction method [5], which has a parameter $\mu \in [0,1] \in \mathbf{R}$ that determines whether the networks generated are SW ($0<\mu<<1$) or SF ($\mu=1$) or a "hybrid" network topology part-way between SW and SF ($0<\mu<1$). We find that the best scores for system-level performance metrics of the simulated data-centres are given by "hybrid" values of $\mu$ significantly different from pure-SW or pure-SF.

The structure of this paper is as follows. In Section II we give further details of *SPECI*, sufficient for the reader to comprehend the new results presented in this paper. In Section III we validate our modifications to KE that turned it into a directed-graph generator. In Section IV, we summarise some of the results presented in [1] against which we then compare the outputs from the directed-KE hybrid complex subscription network. Our conclusions are given in Section V.

## 2  Explanation of SPECI

A software layer (so-called "middleware") that is responsible for job scheduling, load-balancing, security, virtual network provisioning, and resilience binds the components of the DCs together, and is the DC's management layer. Scalability requires the performance not to rapidly degrade as the number of components increases, so that it remains feasible to operate in the desired size range [8]. Yet it is unlikely that all properties in middleware will scale linearly when scaling up the size of DCs.

Because the middleware's settings and available resources change very frequently, it needs to continuously communicate new policies to the nodes. Traditionally middleware manages its constituent nodes using central control nodes, but hierarchical designs scale poorly. Distributed systems management suggests controlling large DCs using policies that can be broken into components for distribution via peer-to-peer (P2P) communication channels, and executed locally at each node. P2P solutions scale better, but can suffer from problems of timeliness (how quickly updated policies will be available at every node) and of consistency (whether the same policies are available and in place everywhere).

The simulation contains a number ($n$) of nodes or services connected through a network. Each of these nodes can be functioning (alive) or not, and state changes occur at a change-rate $f$. To discover the aliveness of other nodes, each node provides an arbitrary state to which other nodes can listen. When the state can be retrieved the node is alive, otherwise it is not. Every node is interested in the aliveness of some of the other nodes, the value of "some" being a variable across all nodes. Each node maintains a subscription list of nodes in whose aliveness it is interested. *SPECI* provides a monitoring probe of the current number of inconsistencies, and the number of network packets dealt with by every node, per unit time (here every second). For now, the simulator assumes uniform costs for connecting to other nodes, but we intend to explore varying the connection costs in a meaningful way, in future work.

Initially, we observed the number of nodes that have an inconsistent view of the system. A node has an inconsistent view if any of the subscriptions that node has contains incorrect aliveness information. We measure this as the number of inconsistent nodes, observed here once per $\Delta t$ (=1sec). After an individual failure or change occurs, there are as many inconsistencies as there are nodes subscribed to the failed node. Some of these will regain a consistent view within $\Delta t$, i.e. before the following observation, and the remaining ones will be counted as inconsistent at this observation point. If the recovery is quicker than the time to the next failure, at the subsequent observations fewer nodes will be inconsistent, until the number drops to zero. When the update retrieval method requires aliveness data to be passed on, more hops would make us expect more inconsistencies, as out-dated data could be passed on. This probing was carried out while running SPECI with increasing failure rates and scale, and using each of these combinations with each of our four protocols. We scaled $n$ though DC sizes of $10^2$, $10^3$, $10^4$ and $10^5$ nodes. We assume that the number of subscriptions grows slower than the number of nodes in a DC, and so we set the number of subscriptions to $n^{0.5}$ per node.

For each of these sizes a failure or change rate distribution $f$ was chosen such that on average in every minute 0.01%, 0.1%, 1%, and 10% of the nodes would fail. A gamma distribution was used with coefficients that would result in the desired rate of failures. Each pair of configurations was tested over 10 independent runs, each lasting 3600 simulation time seconds, and the average number of inconsistencies along with its standard deviation, maximum, and minimum number were observed. The half-width of the 95% confidence intervals (95% CI) was then calculated using the Student's $t$-distribution for small or incomplete data sets.

Formally, the transitive P2P protocol is represented by a directed acyclic graph (DAG) with these distribution topologies. State-changes are introduced to this DAG at a rate $f$, and load and inconsistencies during the propagation of the state changes are measured. The work reported here continues investigating the trade-off between load and inconsistencies between SW and SF, and continues exploring further subscription topologies by investigating hybrid complex network topologies, and show benefits such hybrid topologies can bring as distribution patterns for services in large-scale data centres. The following subsection explains the complex network topologies used.

### 2.3 Complex Network Topologies

The original algorithms for constructing these styles of network are for undirected graphs; but the *SPECI* subscription network requires a directed graph. Here we describe the directed implementations used as well as the applicability of the topology.

**Small world (SW)** undirected networks were first constructed by Watts and Strogatz using an algorithm described as rewiring a ring lattice [3]. Kleinberg has proposed a directed version of SW networks, which starts from a two-dimensional grid rather than a ring in his work searching for a decentralised algorithm to find shortest paths in directed SW [9]. However, here we stay closer to the original algorithm by Watts and Strogatz with the only modification that the initial condition is a ring lattice with outgoing directed edges to all $k$ nearest neighbours, $k$ being the number of subscriptions in our model. Thus, between neighbouring edges there will be two edges, one in each direction. As rewiring probability $p$=0.1was used, a value Watts and Strogatz showed to be large enough to exhibit short path lengths, and at the same time small enough to exhibit a high transitivity of the network [3]. SW networks have a high clustering coefficient or transitivity in the directed case (i.e. many of the neighbours are themselves neighbours) similar to regular lattice graphs, but at the same time they have low diameters (i.e. short average path lengths) as found in random graphs. Watts and Strogatz [3] showed that networks with such properties arise naturally in many fields and are commonly found in natural phenomena. In a data centre, we could imagine the subscription graph having a SW distribution when components are initially placed close to each other, and over time change their location or their functionality (e.g. as a result of load-balancing) thereby turning from local into long range contacts.

**Scale-free (SF)** undirected networks were first constructed by Barabási and Albert (BA) by growing a network with preferential attachment [4, 10]. Unfortunately, the Barabási-Albert model cannot be directly transferred into a directed network. Several attempts exist to model a directed version. They mostly differ in how the initial network that is to be grown is generated, and in how the problem is overcome that directed SF networks are often acyclic, unlike real world networks. For a discussion of weaknesses of directed versions of the Barabási Albert model see the Newman's review of complex networks [11]. For our work, the implementation proposed by Yuan and Wang [12] is used. This starts with a fully connected network, and then grows it using preferential attachment. This results in a citation-network with feed-forward characteristics, where each new node is connected pointing towards the existing nodes. However, with a small probability $p$ the direction of edges is inverted, to allow the generation of cycles. We used $p=0.15$, which was a value under which Yuan and Wang found the network to be "at the edge of chaos" when transformed into Kauffman's NK model [13]. SF networks have low diameters, and been found in many complex networks in the real world, too. For instance, many aspects of the internet are SF networks. SF networks could have applicability for subscriptions when there are components with different importance for other nodes, for instance.

**Hybrid complex networks** have all three characteristics of the previous complex networks, namely low diameters, a high clustering coefficient, and a power-law distribution, from which the SW and SF each exhibit only two but lack one [5]. There are two algorithms for building hybrid complex networks: The one by Klemm and Eguiluz (KE) [5] is used in this work, and there is another implementation by Holme and Kim [14] that is not studied here. The KE model starts with a fully connected network of $k$ nodes, which are all marked as "active", and is grown by iteratively adding nodes that connect to all active nodes and then take the active token from one of the previously active nodes, where the probability of deactivation is proportional to the invers of its current degree. However, with probability $\mu \in [0,1] \in \mathbb{R}$ each of the introduced edges becomes a long-range connection that is connected using plain preferential attachment as in the BA model. Thus, for $\mu=1$ the model is identical to BA, for $\mu=0$ it is a network with high clustering and power-law degree distribution but without low diameter, and for "hybrid" values part-way between ($0<\mu<1$), all three characteristics are present and the cross-over between the models can be studied.

The original KE implementation is an undirected model, as was the case with BA. We applied the same method Yuan and Wang [12] used for plain BA to create directed networks. A directed feed-forward graph of the KE model is grown, and an "invert probability" determined the likelihood that some of the nodes would be inverted to allow the creation of cycles in the graph. Section 3 shows that the characteristics of the KE are still valid after this modification, with transitivity replacing clustering coefficient observations.

## 3 Growing directed scale-free small-worlds

In Subsection 2.4 the KE algorithm to generate hybrid complex networks was described. In this section we show that the directed version we use has the same properties as the original KE model, and look at the behaviour when modifying the invert probability. Fig. 1 shows the effect of introducing $\mu \ll 1$ of random links to the highly clustered model. The average path length drops rapidly with the introduction of these links, while the transitivity is preserved until $\mu$ reaches the order of 1. The graph in Fig. 1 has the same shape as the graph in the original undirected version of KE, i.e. Fig. 1 in [5]. Note that values for transitivity replace the values for clustering coefficient, as clustering is not defined for directed networks. Fig. 2 shows the effect of varying the invert probability for generating the directed version of the graph. Transitivity and Path length appear to change at similar rates throughout the values for the invert parameter.

## 4 Results

The results from [1 and 2] showed that the distribution topology of collaborative components in a data-centre has an impact on the performance of the middleware, and needs to be taken into account when tuning and selecting the topology of the middleware-dependency network. This motivates us to search for preferable distribution topologies. It also demonstrates the value of rigorous simulation tools that allow exploratory studies of planned designs for large-scale data centres, identifying effects that were not anticipated or deliberately accounted for at design phase. We found that the transitivity in the small-world subscriptions caused benefits in the load generated, but also caused the existence of many more inconsistent nodes, as out of date information was passed on. The use of the KE model throws further light on the influence of path lengths and transitivity to the middleware performance. Fig. 3 shows the outcome of *SPECI* runs where the subscriptions are distributed according to a KE network: The level of inconsistencies in the graphs with larger $n$ and $f$ drops with increasing $\mu$ up to a certain key value $\mu^*$; after this threshold value the inconsistencies rise sharply. Note that the value for $\mu=1$ differs from the value in the previous section, because the algorithm here starts with an initial set of $k$ connected nodes, while the previously used implementation of BA started with an initial set of *(k-1)* unconnected nodes which were then connected to and from the $k$-th element.

This outcome leads to two assumptions: First, the best scores for system level performance occur under hybrid topologies that are neither entirely SW nor entirely SF. Second, the inconsistencies fall as the path lengths drop. When $\mu$ grows to the extent that the transitivity drops, a sharp rise in inconsistencies can be observed. This contradicts our previous belief that the transitivity caused the higher levels of inconsistencies in our previous results from plain SW. Based on these insights we can recommend the subscription topology in a data centre to have both high transitivity and low path lengths.

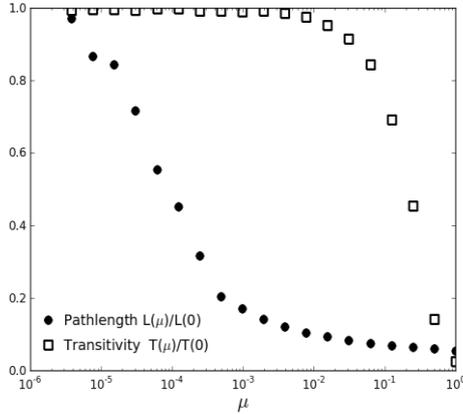
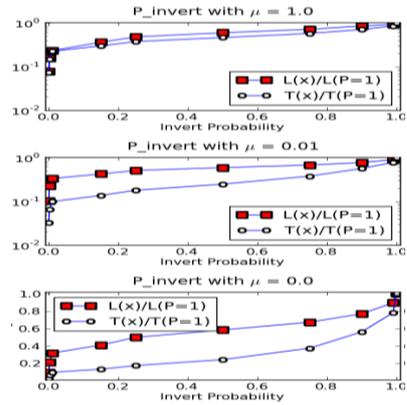

**Fig. 1.** Small-world effect in directed scale-free networks. Tuning the factor μ in the directed version of KE shows the same effects to Path length and Transitivity as the undirected version [5] does to path length and clustering coefficient. n=$10^4$ nodes with average out-degree k=20 where used, as in Fig. 1 of the original KE publication [5], and the invert probability was set to 0.15.

**Fig. 2.** Unlike the factor μ, the invert probability used to create the directed model does not exhibit comparable effects on Path length and Transitivity. In BA (μ=1) Yuang and Wang reported an invert Probability around 0.15 to reflect common scale-free networks

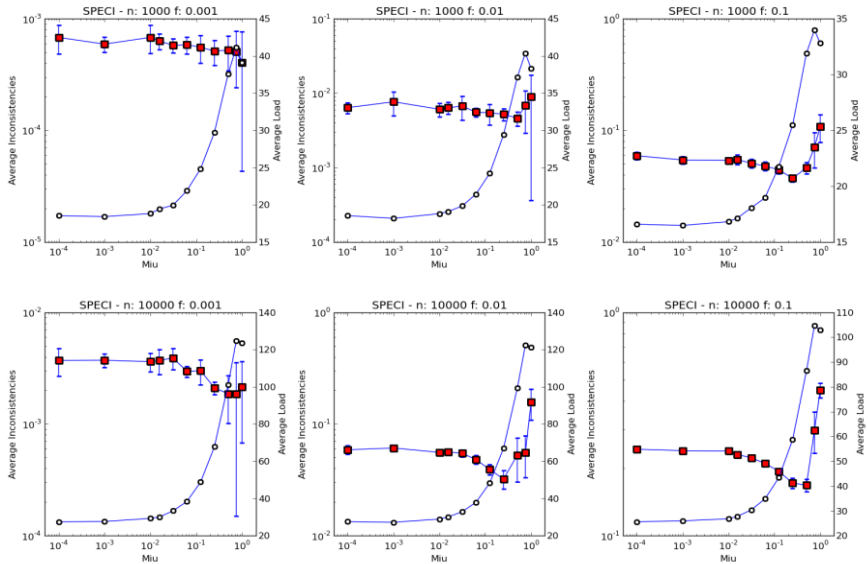

**Fig. 3.** Increasing $\mu$ from small-worlds ($\mu$=0) to scale-free networks ($\mu$=1) for DC sizes of $10^3$ and $10^4$ with change rates of 0.1%, 1% and 10%. The load increases for larger $\mu$. The inconsistencies, however, drop until $\mu$=0.2, and rise sharply for larger $\mu$. This shows that rather than a pure BA distribution of subscriptions in a data centre, a hybrid model would be desirable.

## 5  Conclusion

In this paper we have demonstrated the use of rigorous simulation tools to allow exploration of planned designs for large-scale data centres. Such simulations can be used to identify (and hence avoid) unanticipated and undesirable behaviours. Our results presented here indicate that *hybrid* complex network topologies of dependencies between the computing components may bring benefits. The best scores for global performance levels were found for values that were neither purely SW nor BA SF networks. Finally, the KE model has helped to identify the level of contribution of path length and transitivity to the results in our previous results.